\documentclass[10pt,letterpaper]{article}
\usepackage[top=0.85in,left=2.75in,footskip=0.75in]{geometry}

\usepackage{amsmath,amssymb}

\usepackage{changepage}

\usepackage{textcomp,marvosym}

\usepackage{cite}

\usepackage{nameref,hyperref}

\usepackage[right]{lineno}

\usepackage[nopatch=eqnum]{microtype}
\DisableLigatures[f]{encoding = *, family = * }

\usepackage[table]{xcolor}

\usepackage{array}

\usepackage{booktabs}
\usepackage{graphicx}
\usepackage{multirow}

\newcolumntype{+}{!{\vrule width 2pt}}

\newlength\savedwidth

\raggedright
\usepackage[aboveskip=1pt,labelfont=bf,labelsep=period,justification=raggedright,singlelinecheck=off]{caption}

\makeatletter
\renewcommand{\@biblabel}[1]{\quad#1.}
\makeatother

\usepackage{lastpage,fancyhdr,graphicx}
\usepackage{epstopdf}
\fancyheadoffset[L]{2.25in}
\fancyfootoffset[L]{2.25in}
\begin{document}
\vspace*{0.2in}

\begin{flushleft}
{\Large
\textbf\newline{Use as Directed? A Comparison of Software Tools Intended to Check Rigor and Transparency of Published Work}
}
\newline
\\
Peter Eckmann\textsuperscript{1},
Adrian Barnett\textsuperscript{2},  
Alexandra Bannach-Brown\textsuperscript{3},  
Elisa Pilar Bascunan Atria\textsuperscript{3},  
Guillaume Cabanac\textsuperscript{6},  
Louise Delwen Owen Franzen\textsuperscript{3},  
Małgorzata Anna Gazda\textsuperscript{9},  
Kaitlyn Hair\textsuperscript{8},  
James Howison\textsuperscript{11},  
Halil Kilicoglu\textsuperscript{7},  
Cyril Labbe\textsuperscript{4},  
Sarah McCann\textsuperscript{3},  
Vladislav Nachev\textsuperscript{3},  
Martijn Roelandse\textsuperscript{10},  
Maia Salholz-Hillel\textsuperscript{3},  
Robert Schulz\textsuperscript{3},  
Gerben ter Riet\textsuperscript{5},  
Colby Vorland\textsuperscript{12},
Anita Bandrowski*\textsuperscript{13,14},
Tracey Weissgerber*\textsuperscript{3,15,16}
\\
\bigskip
\textbf{1} Department of Computer Science and Engineering, UC San Diego, La Jolla, CA, United States\\
\textbf{2} School of Public Health and Social Work, Queensland University of Technology, Kelvin Grove, Australia\\
\textbf{3} QUEST Center for Responsible Research, Berlin Institute of Health at Charité Universitätsmedizin Berlin, Germany\\
\textbf{4} Université Grenoble Alpes, France\\
\textbf{5} Hogeschool van Amsterdam, Amsterdam University of Applied Sciences, Amsterdam, Netherlands\\
\textbf{6} Université de Toulouse \& Institut Universitaire de France, France\\
\textbf{7} School of Information Sciences, University of Illinois Urbana-Champaign, IL, United States\\
\textbf{8} UCL Social Research Institute, University College London, London\\
\textbf{9} Department of Biological Sciences, University of Montréal,1375 Avenue Thérèse-Lavoie-Roux, H3C 3J7 Montréal, Québec, Canada\\
\textbf{10} martijnroelandse.dev, Ouderkerk aan de Amstel, Netherlands\\
\textbf{11} Information School of the University of Texas at Austin, Austin, TX, United States\\
\textbf{12} Indiana University School of Public Health-Bloomington, IN, United States\\
\textbf{13} Department of Neuroscience, UC San Diego, La Jolla, CA, United States\\
\textbf{14} SciCrunch Inc.\\
\textbf{15} CIBB, Center for Innovative Biomedicine and Biotechnology, University of Coimbra, Coimbra, Portugal\\
\textbf{16} CNC-UC, Center for Neuroscience and Cell Biology, University of Coimbra, Coimbra, Portugal\\
\bigskip

% Use the asterisk to denote corresponding authorship and provide email address in note below.
* abandrowski@ucsd.edu\\
* tracey.weissgerber@uc.pt

\end{flushleft}
\section*{Abstract}
The causes of the reproducibility crisis include lack of standardization and transparency in scientific reporting. Checklists such as ARRIVE and CONSORT seek to improve transparency, but they are not always followed by authors and peer review often fails to identify missing items. To address these issues, there are several automated tools that have been designed to check different rigor criteria. We have conducted a broad comparison of 11 automated tools across 9 different rigor criteria from the ScreenIT group. We found some criteria, including detecting open data, where the combination of tools showed a clear winner, a tool which performed much better than other tools. In other cases, including detection of inclusion and exclusion criteria, the combination of tools exceeded the performance of any one tool. We also identified key areas where tool developers should focus their effort to make their tool maximally useful. We conclude with a set of insights and recommendations for stakeholders in the development of rigor and transparency detection tools. The code and data for the study is available at \url{https://github.com/PeterEckmann1/tool-comparison}.

% Please keep the Author Summary between 150 and 200 words
% Use first person. PLOS ONE authors please skip this step. 
% Author Summary not valid for PLOS ONE submissions.   
%\section*{Author summary}
%Lorem ipsum dolor sit amet, consectetur adipiscing elit. Curabitur eget porta erat. Morbi consectetur est vel gravida pretium. Suspendisse ut dui eu ante cursus gravida non sed sem. Nullam sapien tellus, commodo id velit id, eleifend volutpat quam. Phasellus mauris velit, dapibus finibus elementum vel, pulvinar non tellus. Nunc pellentesque pretium diam, quis maximus dolor faucibus id. Nunc convallis sodales ante, ut ullamcorper est egestas vitae. Nam sit amet enim ultrices, ultrices elit pulvinar, volutpat risus.

%\linenumbers

\section*{Introduction}
The reproducibility crisis remains a central concern \cite{baker20161, diaba2021reproducibility} in scientific fields ranging from psychology \cite{camerer2018evaluating} to cancer biology \cite{errington2021challenges}. The causes of this crisis include lack of standardization and transparency in scientific reporting \cite{korbmacher2023replication}, which has led to the addition of various checklists and instructions for grantees and authors. Popular checklists such as the CONSORT \cite{hopewell2025consort} or ARRIVE \cite{percie2020arrive} guidelines for human and preclinical animal studies, respectively, have been proposed and added to prominent journals’ instructions to authors. Funders such as the NIH \cite{nih_rigor_repro_guidance_2024} have also implemented checklists as part of the grant submission process. Checklists may increase awareness of issues affecting reproducibility, but evidence suggests they do not significantly improve reporting quality \cite{leung2018arrive, blanco2019scoping}.

While reproducibility consists of many factors, perhaps one of the easiest to address is transparency. Transparent, high quality reporting is achievable even if a study is already completed, but was not conducted in a fully rigorous manner. This includes details of the methods used such as blinding, research materials, and placing data and software into locations that are as open as possible and as closed as necessary \cite{martone2024past, nature_six_factors_reproducibility_2019}. Peer review can help to address many of these issues, but human reviewers often fail to point out important missing information such as catalog numbers for key resources, and may not comment when a criterion like blinding is absent when it is not commonly reported in the field \cite{horbach2019ability}. Humans are also less likely to flag some problematic practices like plagiarism, which requires searching through millions of documents, and manipulated figures, which requires extensive time, training and experience. Therefore, detecting these practices is best done through automated analysis. Paper mills produce problematic papers at scale, which can overwhelm the traditional peer review system and is contributing to an unprecedented number of retractions \cite{oransky_retractionwatch_2022}. For these reasons, the use of software tools has been proposed to automatically flag missing criteria critical for transparency \cite{schulz2022future}. Many publishers already use central platforms like the STM Integrity Hub \cite{van-rossum_research_integrity_hub_2024} to check submitted papers for fraudulent or suspicious content using automated tools, which would otherwise be difficult to catch via traditional peer review. Other tools employ image and text mining techniques to search for criteria such as blinding, power calculations, randomization, open sourcing of code and data in manuscripts, and checking figure quality \cite{menke2022establishing, riedel2020oddpub, wang2022risk, kilicoglu2021toward}.

\begin{figure}
\begin{center}
\centerline{\includegraphics[width=\textwidth]{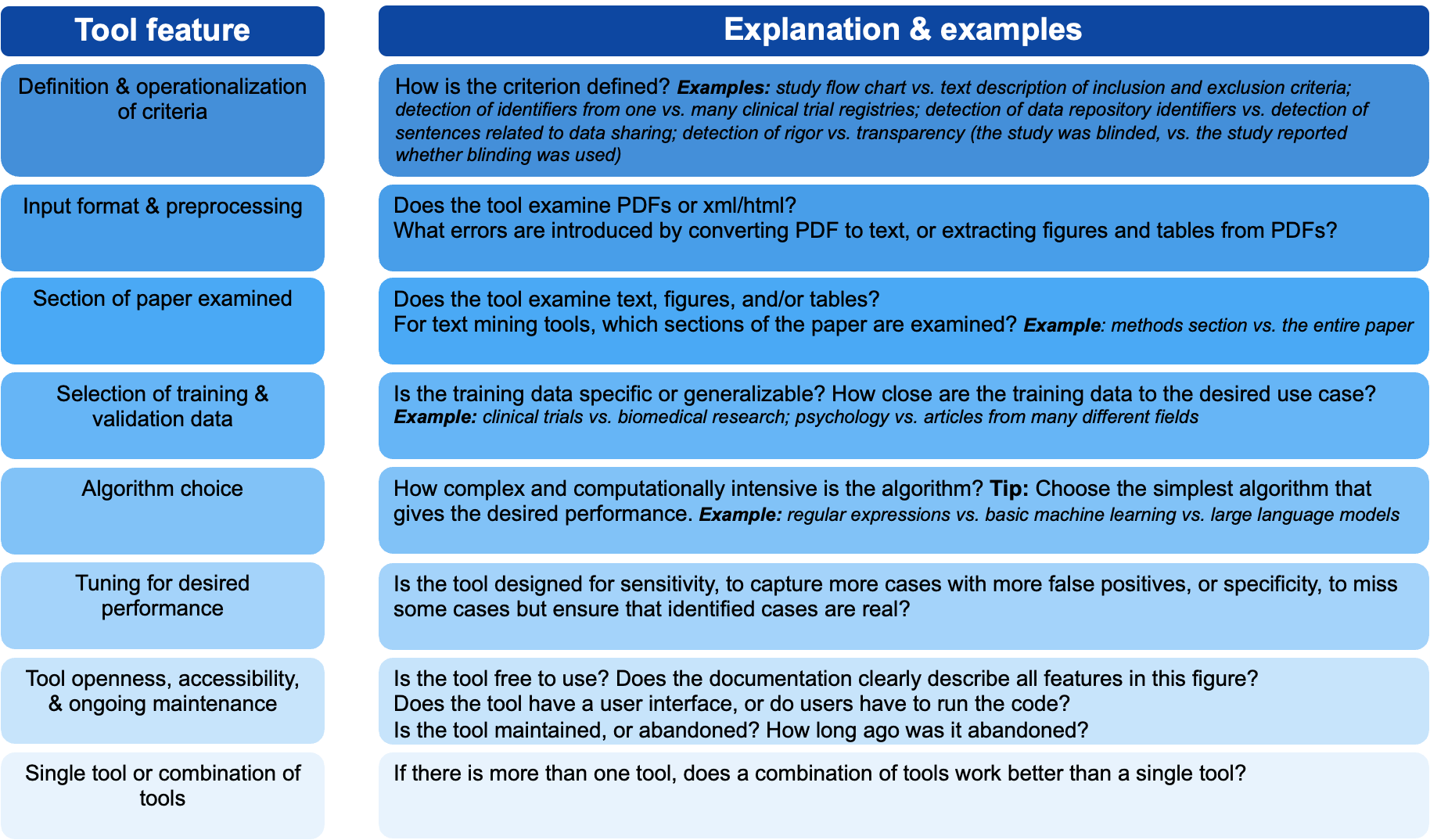}}
\caption{\textbf{Design features that may affect tool performance.} This figure highlights and explains eight features that can affect tool performance,and provides explanations and guiding questions for each feature.
}
\label{intro-fig}
\end{center}
\end{figure}

With the expansion of automated tools available to detect rigor criteria, there is an absence of work comparing the efficacy of the tools. Selecting a tool for a given use case can be difficult, as many tools purport to search for the same or similar things. Direct comparisons would help tool developers and users to understand the design and performance differences between tools (examples given in Figure \ref{intro-fig}). Comparisons would also help users, including publishers, reviewers, and metascientists, to move beyond selecting the tool that performs best in one context, and consider which tool, or combination of tools, fits best with the user's intended purpose. A comparison would further benefit tool developers to determine which types of tools or which combination of tools are most effective in solving a particular problem, discover areas where existing tools are insufficient, and determine what design decisions affect tool performance.

Most papers that introduce new tools compare performance against similar tools, but the dataset used for these comparisons is not standardized. In the worst case, some tool developers may try multiple datasets and pick the one where their tool has the highest performance, making it even more difficult to assess the tool’s true performance \cite{semmelrock2023reproducibility}. These practices can lead to “phantom progress,” where each new paper claims to reach state-of-the-art performance, but independent evaluation reveals no differences between old and new methods (e.g. \cite{ferrari2019we}). Thus, a broad, independent comparison of rigor criteria tools is needed.

We seek to address the need by performing a broad, independent comparison across multiple tools and rigor criteria. We built a dataset encompassing a random subset of 1,500 open access manuscripts in PubMed Central, to ensure that our results would be applicable to many biomedical fields. We evaluated 9 rigor criteria, and ran a suite of 11 tools on all papers in our dataset. Human curators assisted in labeling a gold standard dataset for each criterion, primarily focusing on cases where the tools disagreed, which we used to compute performance characteristics for each tool. After applying this standardized approach to many criteria and tools, we use our comparisons to examine progress in tools for detecting rigor criteria and to provide insights for tool developers and users.

\section*{Methods}
The tools examined here were created by members of ScreenIT, a community of curators, developers, and other scientists that focus on the problem of scientific rigor and reproducibility \cite{weissgerber2021automated}. The goal of this community is to alleviate current problems in poor research quality by the independent development of rigor-checking tools. In this study, we analyzed which tools were more and less useful for various tasks. We attempted to determine which tool performed better, but also considered whether a combination of tools was better suited to address a broader problem in reproducibility. Not all tools are directly comparable because they define the presence or absence of criteria differently; indeed, none of the tools define even seemingly simple things like the presence of code in the same way. For example, some tools assess the availability of open code, while other tools simply detect a statement about code sharing, such as “code is not available for this study.” The sections below contain an overview of the methods used, but see the Supplementary Information for specific methods for each rigor criteria.

\subsection*{Paper extraction}
XML files for 1,500 papers from PubMedCentral’s Non-Commercial Open Access Subset (available at \url{https://ftp.ncbi.nlm.nih.gov/pub/pmc/oa_bulk/oa_noncomm/}) with a PMCID starting with PMC008 (approximately March 2021 to April 2022) were selected using a simple random sample. For each paper, both full text (all text contained within the XML’s \texttt{<body>} tag) and the methods section (all text within sections with titles that contain the words “method” or “procedure”) were extracted. Two tools screen PDFs to evaluate figures and tables; therefore, PDFs of all 1,500 papers were also retrieved automatically.

\subsection*{Test set construction}
In general, for each comparison (other than tools that scanned the PDF version of the manuscript), we ran all tools on the extracted full text. In cases where all tools agreed on the presence or absence of the criteria of interest, we treated that prediction as the “gold standard” result in our test set and performed no manual labeling. All papers where at least one of the tools disagreed on the presence or absence of a given criteria were examined by a curator assigned to that comparison task. This technique is similar to that of the Cranfield methodology \cite{cleverdon1960aslib, voorhees2007trec} from information retrieval, where it is impossible to manually curate every document; therefore, one has to use the tools being studied to extract a smaller set of relevant documents to curate.
Human curators were blinded to which tool(s) disagreed. This was done to reduce bias as some curators were developers of the tools being analyzed. In order to speed curation time, we showed the human curator the extracted sentence (e.g. “Code for this paper is available at …”) so that curators could quickly identify positive cases. To reduce bias, we randomly selected a tool and displayed its extracted sentence, if it existed.

We sought to estimate the rate of error in our test set where all tools agreed. While it would be infeasible to manually curate every paper in the test set, we extracted 100 papers out of our test set for each rigor criteria where all tools agreed on the classification, and manually classified each extracted paper. This was to estimate the rate at which all tools were wrong, so that we could more accurately report performance characteristics of the tools. If we had at least 50 positive (all tools agreed that the criteria was present) and at least 50 negative examples (all tools agreed that the criteria was absent), then the set consisted of 50 positives and 50 negatives. If we had fewer than 50 positive or negative papers, we used more of the other type to reach 100 papers.

\subsection*{Generic definitions of items}
The definition of a criterion’s presence or absence frequently differed between tools; therefore, curators were instructed to use their best judgment as neutral observers, and not the exact definition that was used by one of the tool makers. This assessment approach was designed to ensure that the criteria were not dependent on any given tool. For example, there are specific definitions of a software tool that were strictly enforced in the training data for SoftCite (Du et al, 2021). These specific definitions were not provided to the curator, who instead used their best guess as to whether the item was or was not a piece of software (more specifics in each analysis subsection). This decision was made to increase the generalizability of the results, but it also introduces a type of bias. Curators tracked their definitions and processes in notes, which are summarized below for each of the analyses. 

\subsection*{Ensemble model}
To examine the potential value of combining the tools, a logistic regression ensemble model was trained and tested on the 1,500 paper set for each rigor criteria. The ensemble model was trained to make a classification based on a linear combination of the binary results of the individual tools. Sklearn’s (RRID:SCR\_019053) LogisticRegression, trained with cross-entropy loss, was used for this task. Overfitting on the training set was not a concern due to the low expressiveness of a linear model with only four (or less, depending on the number of tools) parameters. Nonetheless, we assessed overfitting using the method described below in “statistical tests”.

\subsection*{Statistical tests}
Each tool was assessed against the test set using an adjusted accuracy, precision, recall, and F1 score. This adjustment was made to reduce inaccuracies introduced by the test set construction method. We used the results from the manual curation of cases where all tools agreed to compute a “presumed positive rate” (PPR), which is the rate at which papers presumed to be positive were actually positive, and a “presumed negative rate” (PNR), the rate at which papers presumed to be negative were actually negative. We used these rates to compute the following adjusted values:

\begin{align*}
    adjTP &= TP * PPR\\
    adjFP &= FP * PNR\\
    adjFN &= FN * PPR\\
    adjTN &= TN * PNR
\end{align*}

where TP, FP, FN, and TN are the number of true positive, false positive, false negative, and true negative examples, respectively.

Therefore, the adjusted values will always be more conservative than the unadjusted. The adjusted accuracy, precision (also known as the “positive predictive value”), recall (also known as the “sensitivity”), and F1 score (the harmonic mean of precision and recall) were then calculated as:

\begin{align*}
accuracy &= \frac{adjTP+adjTN}{adjTP+adjTN+adjFP+adjFN}\\
precision &= \frac{adjTP}{adjTP+adjFP}\\
recall &= \frac{adjTP}{adjTP+adjFN}\\
F1 &= \frac{2}{1/precision + 1/recall}
\end{align*}

For each ensemble model that we trained for a given rigor criterion, we report what we call the “Function learned” to make the model’s learned decision process transparent. This was done by first inputting all permutations of different binary tool results to the model to generate a truth table. Then, we performed boolean simplification to generate a simplified expression of the function learned by the ensemble.

To assess overfitting of the ensemble models, we trained the ensemble model on random 80\% subsets of the training data (80\% of the original dataset randomly sampled without replacement), and evaluated which percent of trained models had the same decision function as the model trained on all data. A high percentage indicates that overfitting is unlikely, because the model learns the same function despite the underlying data being different subsets of the larger dataset. Note that while the parameters might be slightly different, the models trained on the subset of data that have the same “Function learned” will have identical test-time behavior.

For comparisons between tools, a two-tailed t-test with variances calculated from sample 
proportions was used to compare the classification accuracies \cite{sullivan1992robustness}. Additionally, Gwet’s agreement among each pair of tools, and among each tool to the true results, was calculated using the irrCAC python package (RRID:SCR\_023176). Gwet’s statistic measures the agreement between raters whilst adjusting for chance agreement \cite{wongpakaran2013comparison}. We did not take into account the presumed positive/negative rate when calculating Gwet’s agreement.

\section*{Results}
A summary of the rigor criteria and tools we analyzed is presented in Table \ref{restable1}, and a summary of the most important differences between tools is presented in Table \ref{restable2}. For readability, we summarize the findings from each tool comparison in the below sections. The Supplementary Information contains details for each comparison, as well as performance statistics.

\begin{table*}
    \caption{\textbf{Overview of individual comparisons}. We highlight in overview form each tool comparison that we performed. A dash (``-'') indicates that the tools differed in one of the following ways, and a cross mark (``X'') denotes that the difference made a substantial imapct on the final performance. Differences include how the definitions of terms are operationalized into a tool (e.g., does the tool recognize power calculations or other means to check for group size), document input format and requirements for preprocessing (extraction of text or images from PDF documents can be challenging and introduce systematic errors), section of the paper examined (related to sensitivity vs specificity as some information may be missed by a tool that does not run on the section of text where the mention appears), selection of training and validation data such as the field (clinical studies vs psychology), algorithm choice (regular expressions vs large language models), the openness \& accessibility of the tool (does the tool have a version with a user interface, is the tool maintained, is the tool free and open code), and the desired performance (the tool may be tuned for sensitivity to capture more cases with more false positives).}
    \label{restable1}
    \begin{center}
    \begin{scriptsize}
    \begin{sc}
    \hspace*{-7cm}
    \addtolength{\tabcolsep}{-0.4em}
    \begin{tabular}{cc|cccccccc}
    \toprule
    & & & & & Selection & & Tool\\
     & Tools & Operationalization & Input format \& & Sections & of training \& & Algorithm &  openness \& & Sensitivity & Ensemble\\
     Criterion & included & of criteria & preprocessing & of article & validation data & choice & accessibility & vs. specificity & useful?\\
     \midrule
     \multirow{3}{*}{Registration} & TRNScreener & \multirow{3}{*}{X} & &\multirow{3}{*}{X}&\multirow{3}{*}{-}& &\multirow{3}{*}{-}&\multirow{3}{*}{X} & N/A\\
     & ctregistries\\
     & SciScore\\
     \midrule
    \multirow{3}{*}{Inclusion \& exclusion criteria} & pre-rob\\ & Barzooka&-&X&-&-&X&-& &Yes\\
    & SciScore\\
    \midrule
    &CONSORT-TM\\
    Blinding & pre-rob& & &-&-&X&-& &No\\
    & SciScore\\
    \midrule
    &CONSORT-TM\\
    Randomization&pre-rob& & &-&X&-&-& &Yes\\
    &SciScore\\
    \midrule
    \multirow{2}{*}{Power calculations}&CONSORT-TM&\multirow{2}{*}{-}& &\multirow{2}{*}{-}&\multirow{2}{*}{-}&\multirow{2}{*}{-}&\multirow{2}{*}{-}& &No\\
    & SciScore\\
    \midrule
    \multirow{2}{*}{Use of software tools}&SoftCite&\multirow{2}{*}{X}& &\multirow{2}{*}{X}&\multirow{2}{*}{X}&\multirow{2}{*}{-}&\multirow{2}{*}{-}& &No\\
    &SciScore\\
    \midrule
    \multirow{2}{*}{Open code}&ODDPub&\multirow{2}{*}{X}& &\multirow{2}{*}{X}&\multirow{2}{*}{X}&\multirow{2}{*}{X}&\multirow{2}{*}{X}&&No\\
    &SciScore \\
    \midrule
    \multirow{2}{*}{Use of problematic cell lines}&PCL Detector&\multirow{2}{*}{X}& &\multirow{2}{*}{X}&\multirow{2}{*}{X}&\multirow{2}{*}{X}&\multirow{2}{*}{X}&\multirow{2}{*}{X}&No\\
    & SciScore\\
    \midrule
    \multirow{2}{*}{Baseline tables}&baseline\_tables&\multirow{2}{*}{X}&\multirow{2}{*}{X}&\multirow{2}{*}{X}&\multirow{2}{*}{X}&\multirow{2}{*}{X}&\multirow{2}{*}{X}&&N/A\\
    & Baseline\\
    \bottomrule
    \end{tabular}
    \end{sc}
    \end{scriptsize}
    \end{center}
\end{table*}

\begin{table*}
    \caption{\textbf{Overview of individual comparisons where tool differences made substantial difference in performance}. A cross mark (“X”) denotes that the highlighted feature was a key determinant of differences in the performance of the tools included in the comparison. Design features that could potentially influence performance were: 1) How the definitions of terms are operationalized into a tool (e.g., does the tool recognize power calculations or other means to check for group size), 2) Document input format and preprocessing requirements (extraction of text or images from PDF documents can be challenging and introduce systematic errors), 3) the section(s) of the paper examined (related to sensitivity vs specificity as some information may be missed by a tool that does not run on the section of text where the mention appears), 4) the selection of training and validation data, such as the field or study type (clinical trials vs preclinical animal studies), 5) algorithm choice (regular expressions vs large language models), 6) the openness \& accessibility of the tool (does the tool have a version with a user interface, is the tool maintained, is the tool free and open code), and the 7) desired performance (the tool may be tuned for sensitivity to capture more cases with more false positives).}
    \label{restable2}
    \begin{center}
    \begin{small}
    \begin{sc}
    \hspace*{-6cm}
    \begin{tabular}{cccccccccc}
    \toprule
    \midrule
    \bottomrule
    \end{tabular}
    \end{sc}
    \end{small}
    \end{center}
\end{table*}

\textbf{Registration:} In the comparison of registration tools, which look for clinical trial numbers and other protocol registration numbers, the definition of what the tool aimed to find was a major source of difference in performance. SciScore’s definition includes protocols (2 were found in the dataset) while other screening tools look for different sets of clinical trial registries. By far the most prevalent registry found in the 1500 papers was clinicaltrials.gov (ctgov, Table \ref{tab:registration-results}) which was responsible for 137 of 169 total items correctly identified. Interestingly, the ctregistries tool, which recognizes the most registries, also accounted for all false positives in this dataset because the various trial registries often share letter number combinations with granting agency numbers, catalogue numbers and medical abbreviations.

\textbf{Inclusions and exclusion criteria:} The comparison of detection of inclusions and exclusions of participants was tested with three tools: pre-rob, SciScore and Barzooka. The first two tools recognize text using an LLM (large language model) and a CRF (conditional random field) while Barzooka recognizes images of flow charts using a CNN (convolutional neural network model). The raw performance of Barzooka was lower than the other two tools, but that was mostly because authors in our set of 1500 papers primarily described inclusions and exclusions using text, not flow charts. However, the performance of the different tools was also highly complementary. When authors described their inclusions using a flow chart, this information was only detected by Barzooka. When inclusions were described in text, this was usually detected by one of the text tools. As a result, the combination of tools was more effective than any individual tool. The ensemble performance accuracy, precision, recall, and the F1 were all in the 0.95 to 0.98 range where the highest performance, F1, of any individual tool was 0.91.

\textbf{Blinding and randomization:} The comparisons for blinding and randomization were both performed using three tools: pre-rob, SciScore and CONSORT-TM (https://pubmed.ncbi.nlm.nih.gov/33647518/). These three tools all processed text using different models with SciScore using CRF, while the other two tools used LLMs, specifically two versions of BERT (https://arxiv.org/abs/1810.04805). For the comparison of blinding, SciScore performed the best (F1 of 0.89) and the ensemble did not add any performance (F1 of 0.89). Detecting randomization was more difficult for the models, and performance was far lower for all tools (F1’s of 0.4 to 0.76), perhaps because “random” is used to refer to other techniques beyond random assignment such as random effect models. CONSORT-TM did relatively poorly, while pre-rob and SciScore performed better. In this case, the training dataset is likely to explain the difference in performance. CONSORT-TM was trained on text from randomized controlled trials. pre-rob was trained primarily on data from preclinical animal studies. SciScore was trained on a very broad dataset with many different types of studies. The ensemble model also outperformed any individual tool (F1 of 0.76), demonstrating the benefits of combining tools.

\textbf{Sample size determination:} In the analysis to detect how sample size was determined (e.g. power or sample size calculations), we tested SciScore and CONSORT-TM. These tools have similar performance, with an F1 of 0.79 and 0.78 respectively. The main difference seemed to be the tuning of the tool. CONSORT-TM was more likely to produce false positives, while SciScore was more likely to produce false negatives. The ensemble model did not improve performance.

\textbf{Software used:} To find mentions of software in papers, we tested SciScore and SoftCite. SoftCite scored a higher accuracy and F1 score (F1 of 0.87 compared to 0.27), and the ensemble learned to just use the SoftCite results. Here the difference was mainly due to a large number of false negatives in the SciScore data, which is likely to occur when software is mentioned outside of the methods section. Furthermore, SciScore was tuned to find or suggest RRID type entities; therefore this tool uses the RRID list of existing software (\url{https://rrid.site/data/source/nlx_144509-1/search}), and does not detect software that isn’t included in this list. The ensemble model did not increase performance over SoftCite alone.

\section*{Discussion}
In this study, we have conducted a broad comparison of 11 automated tools across 9 different rigor criteria. Based on the results of the comparisons, we draw the following conclusions to guide future tool development.

Firstly, while there are overwhelming performance gaps between tools in some comparisons, most comparisons have marginal differences between the tools. In these cases, tool run times, cost, ease of use, and quality/transparency of the output are likely more important factors than the raw performance. Therefore, tool developers should invest effort in tool usability and transparency in addition to pure performance.

Secondly, success when testing a tool on a different dataset than it was trained on is varied. The pre-rob tool was trained on animal studies, and had strong performance in randomization, but poor performance in inclusion/exclusion detection. Therefore, it is important for developers to make sure they train their tool on a dataset very similar to what it will be applied to. Otherwise, they risk substantially worse performance when applied in practice. 

The case of what is and is not a protocol elucidates the differences in tool development that depend on specific definitions as the major source of variability in tool performance. The SciScore definition of “protocols” includes clinical trials but also methods descriptions that one might find in protocols.io, whereas the other tools are only designed to detect clinical trial registrations. Therefore, if the question of a particular investigator is how many protocols are included in studies they might choose SciScore, but if the question is centered on finding clinical trial registrations then a different tool will provide better answers. 

Thirdly, combining tools improves performance over any individual tool in some cases. Combining the best performing tool with tools that did not perform as well led to improved performance, compared to the best tool alone. Therefore, users should not dismiss worse-performing tools that detect similar things to another tool with better performance. If a tool performs detection in a different way or was trained on a different dataset, it is likely that it will still identify things that the tool with the best performance will miss. We also found that combining results from different tools is particularly valuable when a rigor criteria can be expressed in different modalities, like images and text. In these cases, ensembling tools that search different modalities can be extremely helpful to improve performance. Therefore, tool developers should consider what modalities rigor criteria can be expressed in and develop tools for modalities other than text, including multi-modal tools.

The primary limitation of our study is the construction of our gold standard set. Since we used the output of the tools to determine which papers humans would manually curate, it is likely we failed to correctly classify papers that all tools classified incorrectly. We sought to mitigate this problem by manually classifying a random sample of papers where all tools agreed, and using the resulting data to adjust our performance statistics. Other limitations of our study include a limited set of tools and a limited set of rigor criteria tested.

Bulleted lists of insights from our study are provided below for each of the key stakeholders in the development of automated rigor and transparency tools.

\subsection*{Insights for toolmakers}
Toolmakers should consider the following when designing tools, as these decisions can substantially impact performance.

\begin{itemize}
    \item Operationalization of criteria
    \item Input format and preprocessing
    \item Sections of paper examined
    \item Selection of training and validation data
    \item Algorithm choice
    \item Tool openness, transparency and accessibility
    \item Desired performance: Toolmakers can choose to prioritize sensitivity or specificity
\end{itemize}

\subsection*{Insights for new toolmakers using LLMs}
The proliferation of LLMs has prompted many people without coding experience, or other experience developing tools, to start creating tools. In addition to the insights shared for tool makers above, we highlight some specific points for those using LLMs to develop screening tools.

\begin{itemize}
    \item \textbf{Simpler approaches often perform better.} LLMs are inherently complicated, computationally intensive, and very expensive to develop and run. The large energy requirements exacerbate environmental impact. Toolmakers should use the simplest approach possible that achieves the desired performance. Lower complexity reduces computational time, energy needs, costs, and the likelihood of errors in the code. Simpler approaches work especially well for things that are consistently reported in standard ways, in specific sections of the manuscript. The fact that you can use an LLM doesn’t mean that you should.
    \item \textbf{Results may not be reproducible.} LLMs are a black box. Tool developers do not know what criteria they are using to classify, and identical prompts may not give the same response, especially when there are undisclosed version changes. 
    \item \textbf{Tool creators need a stable version that they control.} Otherwise, performance will change each time that the LLM is updated. Tool makers may not know when updates have occurred.
    \item \textbf{Validation is essential.} Many tools are released without any data on performance, which means users have no information on exactly what they are designed to detect, how they were trained, how often they make mistakes/hallucinations, and whether the tool is appropriate for their use case. Validations against human curated gold standard data should be performed regularly, and repeated every time that the tool or underlying LLM is adjusted.
\end{itemize}

\subsection*{Insights for tool users}
\begin{itemize}
    \item Pay attention to tool inputs and preprocessing steps.
    \item Validate the tool on your own dataset to determine whether the tool is appropriate for your use case.
    \item Test ensemble tools. In some cases, one tool is clearly superior. In other cases, a combination of tools performs better than any individual tool.
    \item Know the limitations of the tool(s) that you use.
    \item Consider openness, accessibility, and whether the tool is maintained.
\end{itemize}

\subsection*{Insights for those receiving results from tools}
Those who receive results or reports from tools should pay particular attention to the following details when using and interpreting the reports.

\begin{itemize}
    \item \textbf{Understand the criteria.} Consult documentation in the report to understand how each item is defined, and why the items that are assessed are important. Readers must understand exactly what the tool was designed to detect, and how the criteria were operationalized, to interpret the report. Understanding why the items are important will help users to improve the paper.
    \item \textbf{Some items may not be relevant.} Remember that some items may not apply to every paper. Some tools simply report that the item wasn’t found, without determining whether it was needed. Other tools distinguish between cases where an item is not reported, and cases where the item is not relevant.
    \item \textbf{Expect errors.} When a tool measures many items, it is likely that you will see at least one false positive or false negative in a report. Even among classifiers with very high performance, the likelihood of an error on at least one item is high.
    \item \textbf{Regular users should validate performance.} If you use reports from tools regularly (e.g., editors) or for larger datasets (e.g., metascientists), check some percentage of the reports manually. All tools make mistakes, and regular users should know the types of errors that the tools make and how often these errors occur. This will help you to use the report responsibly.
\end{itemize}

\section*{Acknowledgments}
We would like to thank the members of the ScreenIT group who were not included as authors in this paper, but nonetheless contributed in discussions about this and related projects: Jennifer Byrne, Nicholas Brown, Tim Vines, Thomas Lemberger, Vince Istvan Madai, Julia Menon, Sean Rife, Iain Hrynaszkiewicz, Han Zhuang, Angelo Pezzullo, Andrew Brown, Camila Victoria-Quilla Baselly Heinrich, Rene Bernard, and Bertrand Favier. CL acknowledges the NanoBubbles project that has received Synergy grant funding from the European Research Council (ERC), within the European Union's Horizon 2020 program, grant agreement no. 951393 (\url{https://doi.org/10.3030/951393}). GC received funding from the Institut Universitaire de France (IUF) and the NanoBubbles project that received Synergy grant funding from the European Research Council (ERC), within the European Union’s Horizon 2020 program, grant agreement no. 951393. HK was partially supported by the National Library of Medicine of the National Institutes of Health under the award number R01LM014079. The content is solely the responsibility of the authors and does not necessarily represent the official views of the National Institutes of Health. The funder had no role in considering the study design or in the collection, analysis, interpretation of data, writing of the report, or decision to submit the article for publication. KH is supported by a Wellcome Trust Early-Career Award (306380/Z/23/Z). MR serves as independent contractors for SciCrunch Inc. AB is co-founder and serves as CEO of SciCrunch Inc, a company that works with publishers to improve the scientific literature; the terms of this agreement have been approved by the COI office at the University of California at San Diego. The authors would like to thank the National Institutes of Health [GM144308]. This project has also been made possible in part by grant 2022-250218 from the Chan Zuckerberg Initiative DAF, an advised fund of the Silicon Valley Community Foundation. The funders had no role in study design, data collection and analysis, decision to publish, or preparation of the manuscript.

\nolinenumbers

% Either type in your references using
% \begin{thebibliography}{}
% Text
% \end{thebibliography}
%
% or
%
% Compile your BiBTeX database using our plos2015.bst
% style file and paste the contents of your .bbl file
% here. See http://journals.plos.org/plosone/s/latex for 
\bibliography{main}

\newpage
\section*{Supporting information}
\subsection*{Methods and results for individual criteria}

\subsubsection*{Registration reporting}

\begin{table*}
\caption{\textbf{List of tools for registration reporting and their reported purpose.}}
\label{tab:registration-tools}
\begin{center}
\begin{small}
\begin{sc}
\hspace*{-6cm}
\begin{scriptsize}
\begin{tabular}{c|c|cccc}
\toprule
 &  &  &  &  & Classification\\
Tool category & Tool name & RRID and commit & Searches for & Search in & method\\
\midrule
\multirow{3}{*}{\parbox{3cm}{Clinical trial identification}} & TRNscreener & \parbox{2cm}{RRID:SCR\_019211 (\href{https://github.com/bgcarlisle/TRNscreener/commit/9d02549a8d2ede3995013c8347aa33e0c9220427}{commit})} & \parbox{4cm}{Trial identifier in ClinicalTrials.gov and ISRCTN} & Full text & Regex\\
\cmidrule{2-6}
& ctregistries & \parbox{2cm}{RRID:SCR\_024412 (\href{https://github.com/maia-sh/ctregistries/commit/06c169cfa241ef8feda9e8b78f57b3012c85afd8}{commit})} & \parbox{4cm}{Trial identifier in any WHO ICTRP Primary Registry or Data Provider} & Full text & Regex\\
\cmidrule{2-6}
& ``NCT'' presence & N/A & \parbox{4cm}{Presence of the text “NCT” (to detect trial identifiers in ClinicalTrials.gov, which start with “NCT”)} & Full text & Regex\\
\midrule
\parbox{3cm}{Registration identification (including clinical trials in select registries)} & SciScore & \parbox{2cm}{RRID:SCR\_016251, version 2} & \parbox{4cm}{Trial identifier in ClinicalTrials.gov and EU Clinical Trials Register;
Identifier from Protocols.io, Protocol Exchange, STAR Protocols, JoVE, Bio-protocol, MethodsX, Nature Protocols, Spring Protocols, BioTechniques,and PROSPERO} & Methods section & Regex\\
\bottomrule
\end{tabular}
\end{scriptsize}
\end{sc}
\end{small}
\end{center}
\end{table*}

\begin{table*}
\caption{\textbf{Registration identifiers found in papers.} True positives indicate the number of each type of registration identifier found by any tool and then manually validated. An empty cell indicates that the tool was not built to detect that type of identifier. Registries: \protect \footnotemark}
\label{tab:registration-results}
\begin{center}
\begin{small}
\begin{sc}
%\hspace*{-6cm}
\begin{scriptsize}
\begin{tabular}{c|c|cc|cc|cc|cc}
\toprule
& True positives & \multicolumn{2}{c}{SciScore} & \multicolumn{2}{c}{ctregistries} & \multicolumn{2}{c}{TRNscreener} & \multicolumn{2}{c}{nct}\\
& n & n & \% & n & \& & n & \& & n & \&\\
\midrule
TRN (total) & 169 &55&32.5\%&166&98.2\%&139&82.2\%&137&81.1\%\\
ctgov&137&54&39.4\%&134&97.8\%&136&99.3\%&137&100\%\\
umin&7&&&7&100\%\\
drks&5&&&5&100\%\\
irct&5&&&5&100\%\\
chictr&3&&&3&100\%\\
isrctn&3&&&3&100\%&3&100\%\\
ctri&2&&&2&100\%\\
eudract&2&1&50\%&5&100\%\\
actrn&1&&&5&100\%\\
jrct&1&&&5&100\%\\
kct&1&&&5&100\%\\
ntr&1&&&5&100\%\\
pactr&1&&&5&100\%\\
\midrule
Protocols (total) & 2 & 2 & 100\%\\
FP (total) & 29 & & & 29 & 100\%\\
funding\_id & 13 & & & 13 & 100\%\\
drug\_id &6 & & & 6 & 100\%\\
datapoint & 5 & & & 5 & 100\%\\
catalog\_id & 3 & & & 3 & 100\%\\
medical\_acronym & 1 & & & 1 & 100\%\\
medical\_device & 1 & & & 1 & 100\%\\
\bottomrule
\end{tabular}
\end{scriptsize}
\end{sc}
\end{small}
\end{center}
\end{table*}

\textbf{Tools:} To detect registration reporting, we used the Trial Registration Number screener (TRNscreener; RRID:SCR\_019211; \url{https://github.com/bgcarlisle/TRNscreener/commit/9d02549a8d2ede3995013c8347aa33e0c9220427}), ctregistries (RRID:SCR\_024412; \url{https://github.com/maia-sh/ctregistries/commit/06c169cfa241ef8feda9e8b78f57b3012c85afd8}), SciScore v2 (RRID:SCR\_016251), and PE’s naive tool (Table \ref{tab:registration-tools}). This tool checked for the presence of “NCT”, the prefix for ClinicalTrials.gov identifiers, anywhere in the paper. These tools were applied to the full text extracted from each of the 1500 PMC papers in the test set. All tools searched for trial registration numbers from one or more trial registries, yet the tools differed with regard to which registries were considered, ranging from ClinicalTrials.gov only for the naive ``NCT'' tool, to all registries in the World Health Organization (WHO) International Clinical Trials Registry Platform (ICTRP) Primary Registries and Data Providers for ctregistries. SciScore considers trial registries akin to protocols, thus additionally finds identifiers for protocol repositories such as protocols.io, which is a case where the definition of “registry” is likely to be a large portion of the difference in performance. In cases where tools extracted multiple identifiers per paper, all identifiers were captured, duplicates were removed, and the remaining unique identifiers were sorted alphabetically to make for easy comparison if tools extracted identifiers in different orders. Duplicate identifiers found across different papers were preserved. This resulted in a dataset of unique identifiers per paper.

\textbf{Manual checks:} One co-author (EB) coded papers, and questions were discussed with two other team members (DLF and MSH). All papers where any tool detected an identifier were manually checked. For each identifier found by any tool, the coder located the unique identifier in the full-text paper; the paper was not checked for further identifiers, and identifiers were not validated (i.e., the registry was not checked for a corresponding registration). Each identifier was classified as a trial registration number, protocol identifier, or false positive. For trial registration numbers and protocol identifiers, the registry or repository of the identifier was captured. For false positives, we noted what the number represented (e.g. funding IDs, medical acronyms, drug IDs). The coder annotated all of the identifier’s locations in the paper, including: abstract, methods or other locations (e.g., introduction, acknowledgments, discussion, etc.). Finally, the coder classified the paper as a research article or another type of paper.

Additionally, a random sample of 50 papers for which no tools found any identifier were manually reviewed in their entirety for the presence of an identifier. The coder scanned the full text of each paper in search of a unique identifier. Any possible identifiers were noted and were reviewed by two additional team members to determine whether any tools should have found the identifier (i.e., false negative). 

We chose not to conduct a performance analysis for this comparison, because differences were relatively minor between tools. Instead, we conducted a descriptive analysis of the differences between the tools, which we believe better captures the relevant differences between the tools.

\footnotetext{ctgov - ClinicalTrials.gov Registry, umin - University Hospital Medical Information Network Registry, drks - German Clinical Trials Register, irct - Iranian Registry of Clinical Trials, chictr - Chinese Clinical Trial Registry, isrctn - ISRCTN Registry, ctri - Clinical Trials Registry-India, eudract - EU Clinical Trials Register, acTRN - Australian New Zealand Clinical Trial Registry, jrct - Japan Registry of Clinical Trials, kct - Korean Clinical Trial Registry, ntr - Dutch Trial Register, pactr - Pan African Clinical Trials Registry.}

\textbf{Descriptive analysis of identifiers:} After running all tools across the 1,500 paper test set, a total of 199 hits for unique registration identifiers unique per paper were found by one or more tools in 117 papers. TRNscreener found 139 hits across 72 (4.8\%) papers, ctregistries 195 across 113 (7.5\%), SciScore 57 across 43 (2.9\%), and “NCT” was present 137 times across 67 papers (4.5\%). Results are summarized in Table \ref{tab:registration-results}.

In our manual checks of the 200 identifiers, we found that 169 were trial registration numbers (TRN), 2 were protocol identifiers, and 29 were false positives (e.g., grant number). The majority of TRN were from ClinicalTrials.gov (137), followed by the Japanese University Hospital Medical Information Network Registry (7), the German Clinical Trials Register (5), and the Iranian Registry of Clinical Trials (5). Additionally, there were nine additional trial registries with fewer than 5 hits per registry.

Taken together, each tool found different types of IDs. SciScore found mainly TRN from ClinicalTrials.gov (n = 54) and the EU Clinical Trials Register (n = 1). It was also the only tool that found two protocol IDs (n = 2). Nct found only TRNs from ClinicalTrials.gov (n = 137). TRNscreener found TRNs from ClinicalTrials.gov (n = 136) and the ISRCTN Registry (n = 3). Finally, ctregistries found TRN from several different registries (n = 166), where the majority were identifiers from ClinicalTrials.gov (n = 134). This tool also found the 29 cases of false positives, where the majority were funding IDs (n = 13).

Finally, it is important to note that there were several papers in the main set that used the registration IDs to cite previous works, mainly in the introduction/background section of the paper. This is relevant as finding an ID by the tools on the papers does not automatically translate to the fact that a study mentioned its own registration id as “good practice”.  For example, in this analysis, 71 out of the 137 ClinicalTrials.gov trial registration numbers were used as reference to other studies. 

Additionally, a manual check was conducted on a set of 50 PMC papers where no registration identifier was found by any of the tools. In this set there was no registration id from the registries mentioned above. However, there were three cases where a different type of registration was discovered: a prospero registry ID (CRD42021225699), a number ID from a registry website (researchregistry7399), and a number ID from the Institutional Animal Care and Use Committee (2014092403). 

More information of this analysis can be found here: \url{https://github.com/delwen/screenit-tool-comparison}.

\subsubsection*{Inclusion/exclusion criteria}
\begin{table*}
\caption{\textbf{List of tools for inclusion/exclusion criteria detection.}}
\label{tab:ie-tools}
\begin{center}
\begin{small}
\begin{sc}
\hspace*{-3cm}
\begin{scriptsize}
\begin{tabular}{c|c|cc}
\toprule
Tool category&Tool name&Searches for&Method of classification\\
\midrule
\multirow{3}{*}{Inclusion/exclusion criteria detection} & pre-rob & \parbox{3cm}{Sentence(s) including inclusion/exclusion criteria} & BERT encoding + linear classifier\\
\cmidrule{2-4}
& SciScore & \parbox{3cm}{Sentence(s) including inclusion/exclusion criteria} & CRFs\\
\cmidrule{2-4}
& Barzooka & Images of flow diagrams & CNN\\
\bottomrule
\end{tabular}
\end{scriptsize}
\end{sc}
\end{small}
\end{center}
\end{table*}

\begin{table*}
\caption{\textbf{Estimated tool performance statistics for inclusion and exclusion criteria detection}. Statistics are estimated based on the gold standard test set of 1,500 papers (see Methods for details). The Ensemble tool uses a combination of the results from all tools to make a classification.}
\label{tab:performance-ie}
\begin{center}
\begin{small}
\begin{sc}
\begin{tabular}{c|cccc}
\toprule
Tool & Accuracy & Precision & Recall & F1\\
\midrule
pre-rob & 0.69 & 0.94 & 0.48 & 0.63 \\ 
SciScore & 0.91 & 0.97 & 0.86 & 0.91 \\ 
Barzooka & 0.61 & \textbf{1.0} & 0.31 & 0.47 \\ 
\midrule 
Ensemble & \textbf{0.96} & 0.98 & \textbf{0.95} & \textbf{0.96} \\ 
\bottomrule
\end{tabular}
\end{sc}
\end{small}
\end{center}
\end{table*}

\begin{table*}
\caption{\textbf{Gwet's agreement among each tool and the gold standard (``true'') for inclusion and exclusion criteria}.}
\label{tab:agreement-inclusion and exclusion criteria}
\begin{center}
\begin{small}
\begin{sc}
\begin{tabular}{ccccccc}
\toprule
 & pre-rob & SciScore & Barzooka & Ensemble & True\\
\midrule
pre-rob & 1.0 & 0.51 & 0.6 & 0.44 & 0.49\\
SciScore &   & 1.0 & 0.38 & 0.91 & 0.84\\
Barzooka &   &   & 1.0 & 0.44 & 0.42\\
Ensemble &   &   &   & 1.0 & 0.92\\
True &   &   &   &   & 1.0\\
\bottomrule
\end{tabular}
\end{sc}
\end{small}
\end{center}
\end{table*}

\textbf{Tools:} The output of three tools used to detect inclusion and exclusion criteria was examined: pre-rob (SCR\_025493), Barzooka (RRID:SCR\_018508) and SciScore (RRID:SCR\_016251) (Table \ref{tab:ie-tools}). Barzooka was run on PDF documents because it detects images, specifically flow diagrams that define the experimental exclusions, whilst the other tools used the full text extracted from XMLs. 

\textbf{Curation:} We performed manual validation of 679 (+100 controls) articles that were previously assessed by both automated screening tools. The human curator for text (AEB) broke up the sentences into three possibilities: Yes, No and ‘it is complicated’. The sentences recognized by the tool were rated on the criteria that they did or did not represent inclusion or exclusion criteria. Sentences that started with “The exclusion criteria were …” or similarly clear statements were not carefully checked but simply accepted as true. Sentences that clearly included study patients or participants, and words that indicated inclusion or exclusion were also marked as “yes”, while sentences that were clearly about excluding or including some experimental variable that was not about the subject population were marked with “no”, for example, “To alleviate artifacts caused by head motion, subjects with mean frame-to-frame displacement > 0.1 mm or maximum frame-to-frame displacement > 0.15 mm were excluded.” In the third category where it was not clear to the curator for 26 sentences whether the exclusion was associated with the subject population or not, for example, “A variable to be analyzed was excluded from the covariates.” In these cases, the curator reviewed the sentence a second time to determine whether the sentence could be clarified. This removed all but 12 borderline statements. For data not directly associated with a particular sentence, the manuscript was examined by first searching for “inclusion” or “exclusion” and then looking for flow diagrams. The human curator for flow charts (TLW) reviewed all flow charts detected by the tool to determine whether they were flow charts depicting the number of included and excluded observations at each stage of the experiment, and explaining reasons for exclusion. 

Positive or ambiguous text statements were pasted into the same document as evidence of the presence of the criterion for documentation. Flow charts were not pasted as they were too big.

\textbf{Results:}\\
\textbf{Function learned: (SciScore OR Barzooka)\\
Percent same with 80\% data splits: 99\%}

Tables \ref{tab:performance-ie} and \ref{tab:agreement-inclusion and exclusion criteria} reports the performance of each tool as well as an ensemble model combining all the tools. While SciScore had a significantly higher accuracy and F1 over pre-rob and Barzooka, the ensemble model was able to further improve the classification performance. Interestingly, the ensemble learned the function (SciScore OR Barzooka). This shows that the performance boost of the ensemble over SciScore is because the ensemble is able to take both image and text modalities into account. These results indicate the importance of ensembling multiple tools together in cases where criteria may appear in multiple modalities, such as text and image.

\subsubsection*{Blinding}
\begin{table*}
\caption{\textbf{List of tools for blinding detection.}}
\label{tab:blinding-tools}
\begin{center}
\begin{small}
\begin{sc}
\hspace*{-3cm}
\begin{scriptsize}
\begin{tabular}{c|c|cc}
\toprule
Tool category&Tool name&Searches for&Method of classification\\
\midrule
\multirow{3}{*}{Blinding} & pre-rob & \parbox{3cm}{Sentence(s) including blinding} & BERT encoding + linear classifier\\
\cmidrule{2-4}
& SciScore & \parbox{3cm}{Sentence(s) including blinding}&CRFs\\
\cmidrule{2-4}
& CONSORT-TM & \parbox{3cm}{Sentence(s) including blinding} & BioBERT-based classifier\\
\bottomrule
\end{tabular}
\end{scriptsize}
\end{sc}
\end{small}
\end{center}
\end{table*}

\begin{table*}
\caption{\textbf{Estimated tool performance statistics for blinding detection}. Statistics are estimated based on the gold standard test set of 1,500 papers (see Methods for details). The Ensemble tool uses a combination of the results from all tools to make a classification.}
\label{tab:performance-blinding}
\begin{center}
\begin{small}
\begin{sc}
\begin{tabular}{c|cccc}
\toprule
Tool & Accuracy & Precision & Recall & F1\\
\midrule
CONSORT-TM & 0.95 & 0.86 & 0.56 & 0.68 \\ 
pre-rob & 0.89 & 0.45 & \textbf{0.87} & 0.59 \\ 
SciScore & \textbf{0.98} & \textbf{0.94} & 0.85 & \textbf{0.89} \\ 
\midrule 
Ensemble & \textbf{0.98} & 0.93 & 0.85 & \textbf{0.89} \\ 
\bottomrule
\end{tabular}
\end{sc}
\end{small}
\end{center}
\end{table*}

\begin{table*}
\caption{\textbf{Gwet's agreement among each tool and the gold standard (``true'') for blinding}.}
\label{tab:agreement-blinding}
\begin{center}
\begin{small}
\begin{sc}
\begin{tabular}{ccccccc}
\toprule
 & CONSORT-TM & pre-rob & SciScore & Ensemble & True\\
\midrule
CONSORT-TM & 1.0 & 0.81 & 0.94 & 0.95 & 0.94\\
pre-rob &   & 1.0 & 0.85 & 0.86 & 0.86\\
SciScore &   &   & 1.0 & 0.99 & 0.98\\
Ensemble &   &   &   & 1.0 & 0.98\\
True &   &   &   &   & 1.0\\
\bottomrule
\end{tabular}
\end{sc}
\end{small}
\end{center}
\end{table*}

\textbf{Tools:} To detect blinding we used three tools, CONSORT-TM (RRID:SCR\_021051), SciScore (RRID:SCR\_016251), and pre-rob tool (RRID:SCR\_025493) (Table \ref{tab:blinding-tools}). All three tools are text based.

\textbf{Curation:} We performed manual validation of 291 (+100 controls) articles that were previously assessed by both automated screening tools. The human curator (KH) determined whether there was any form of blinding present in the study and assigned a “Yes” or “No” category. As blinding is possible at many different stages, the curator also assigned the papers with some form of blinding into four categories to indicate which parts of the study were blinded (blinded conduct - Yes / No / NA, blinded assessment - Yes / No / NA, blinded scoring between investigators - Yes / No / NA), and whether the method of blinding was described (Yes / No). The NA categories were assigned when the form of blinding addressed by the category was not feasible due to the study type. For example, “blinded conduct” is not possible in a retrospective observational study. 

\textbf{Results:}\\
\textbf{Function learned: (CONSORT-TM AND pre-rob) OR (pre-rob AND SciScore) OR (CONSORT-TM AND SciScore)\\
Percent same with 80\% data splits: 95\%}

CONSORT-TM, pre-rob, and SciScore were run on the 1500 paper set to detect blinding (Tables \ref{tab:performance-blinding} and \ref{tab:agreement-blinding}). SciScore showed the best overall performance.  This may be due to the fact that the other LLM-based tools were trained on a more limited dataset and failed to generalize to the broader literature. The ensemble model achieved a similar performance to SciScore, although notably it learned a complex function of the different tools. In this case, it seems that the extra complexity of the ensemble model is not beneficial to prediction. This suggests that ensembling is not always the best approach if the learned function is complex and does not improve prediction by much.

\subsubsection*{Randomization}
\begin{table*}
\caption{\textbf{List of tools for randomization detection.}}
\label{tab:randomization-tools}
\begin{center}
\begin{small}
\begin{sc}
\hspace*{-3cm}
\begin{scriptsize}
\begin{tabular}{c|c|cc}
\toprule
Tool category&Tool name&Searches for&Method of classification\\
\midrule
\multirow{3}{*}{Randomization} & pre-rob & \parbox{3cm}{Sentence(s) including randomization} & BERT encoding + linear classifier\\
\cmidrule{2-4}
& SciScore & \parbox{3cm}{Sentence(s) including randomization}&CRFs\\
\cmidrule{2-4}
& CONSORT-TM & \parbox{3cm}{Sentence(s) including randomization} & BioBERT-based classifier\\
\bottomrule
\end{tabular}
\end{scriptsize}
\end{sc}
\end{small}
\end{center}
\end{table*}

\begin{table*}
\caption{\textbf{Estimated tool performance statistics for randomization detection}. Statistics are estimated based on the gold standard test set of 1,500 papers (see Methods for details). The Ensemble tool uses a combination of the results from all tools to make a classification.}
\label{tab:performance-randomization}
\begin{center}
\begin{small}
\begin{sc}
\begin{tabular}{c|cccc}
\toprule
Tool & Accuracy & Precision & Recall & F1\\
\midrule
CONSORT-TM & 0.87 & 0.43 & 0.37 & 0.4 \\ 
pre-rob & 0.91 & 0.59 & 0.84 & 0.69 \\ 
SciScore & 0.88 & 0.5 & \textbf{0.94} & 0.65 \\ 
\midrule 
Ensemble & \textbf{0.94} & \textbf{0.73} & 0.78 & \textbf{0.76} \\ 
\bottomrule
\end{tabular}
\end{sc}
\end{small}
\end{center}
\end{table*}

\begin{table*}
\caption{\textbf{Gwet's agreement among each tool and the gold standard (``true'') for randomization}.}
\label{tab:agreement-randomization}
\begin{center}
\begin{small}
\begin{sc}
\begin{tabular}{ccccccc}
\toprule
 & CONSORT-TM & pre-rob & SciScore & Ensemble & True\\
\midrule
CONSORT-TM & 1.0 & 0.74 & 0.68 & 0.79 & 0.82\\
pre-rob &   & 1.0 & 0.8 & 0.94 & 0.88\\
SciScore &   &   & 1.0 & 0.86 & 0.83\\
Ensemble &   &   &   & 1.0 & 0.92\\
True &   &   &   &   & 1.0\\
\bottomrule
\end{tabular}
\end{sc}
\end{small}
\end{center}
\end{table*}

\textbf{Tools:} We used the CONSORT-TM (RRID:SCR\_021051), pre-rob (RRID:SCR\_025493), and SciScore (RRID:SCR\_016251) (Table \ref{tab:randomization-tools}) to determine if randomization was present. All of these tools are text based and look for sentences that describe randomization of subjects into groups. 

\textbf{Curation:} We performed manual validation of 461 (+100 control) articles that were previously assessed by both automated screening tools.  The human curators (MAG, AB) assigned sentences into “Yes” and “No” categories. The data were analyzed based on either the sentence extracted by one of the tools, or by manual search using “rando”. The data were classified as “Yes” when randomization was used to split a study cohort into groups, and “No” when there was no randomization or randomization was used for some other purpose (e.g. imputing missing data or cross-validation).

\textbf{Results:}\\
\textbf{Function learned: (pre-rob AND SciScore)\\
Percent same with 80\% data splits: 90\%}
CONSORT-TM, which was trained on randomized controlled trial study data, did relatively poorly, while pre-rob was trained primarily on animal data and SciScore was trained on a very broad dataset, and the animal and broadly trained tools showed a comparably stronger performances (Tables \ref{tab:performance-randomization} and \ref{tab:agreement-randomization}). The ensemble model had a higher accuracy and F1 than any individual tool. In this case, the training dataset is likely to be the key difference for performance on this broad dataset of randomly selected manuscripts. In this case, the ensemble does much better than any individual tool, showing the benefits of the ensemble approach.

\subsubsection*{Power calculations}
\begin{table*}
\caption{\textbf{List of tools for power calculation detection.}}
\label{tab:power-tools}
\begin{center}
\begin{small}
\begin{sc}
\hspace*{-3cm}
\begin{scriptsize}
\begin{tabular}{c|c|cc}
\toprule
Tool category&Tool name&Searches for&Method of classification\\
\midrule
& SciScore & \parbox{3cm}{Sentence(s) including power calculations}&CRFs\\
\cmidrule{2-4}
& CONSORT-TM & \parbox{3cm}{Sentence(s) including power calculations} & BioBERT-based classifier\\
\bottomrule
\end{tabular}
\end{scriptsize}
\end{sc}
\end{small}
\end{center}
\end{table*}

\begin{table*}
\caption{\textbf{Estimated tool performance statistics for power analysis detection}. Statistics are estimated based on the gold standard test set of 1,500 papers (see Methods for details). The Ensemble tool uses a combination of the results from all tools to make a classification.}
\label{tab:performance-power}
\begin{center}
\begin{small}
\begin{sc}
\begin{tabular}{c|cccc}
\toprule
Tool & Accuracy & Precision & Recall & F1\\
\midrule
CONSORT-TM & 0.95 & 0.68 & \textbf{0.96} & \textbf{0.79} \\ 
SciScore & 0.96 & 0.88 & 0.7 & 0.78 \\ 
\midrule 
Ensemble & \textbf{0.97} & \textbf{1.0} & 0.66 & \textbf{0.79} \\ 
\bottomrule
\end{tabular}
\end{sc}
\end{small}
\end{center}
\end{table*}

\begin{table*}
\caption{\textbf{Gwet's agreement among each tool and the gold standard (``true'') for power analysis}.}
\label{tab:agreement-power analysis}
\begin{center}
\begin{small}
\begin{sc}
\begin{tabular}{cccccc}
\toprule
 & CONSORT-TM & SciScore & Ensemble & True\\
\midrule
CONSORT-TM & 1.0 & 0.89 & 0.91 & 0.94\\
SciScore &   & 1.0 & 0.98 & 0.95\\
Ensemble &   &   & 1.0 & 0.96\\
True &   &   &   & 1.0\\
\bottomrule
\end{tabular}
\end{sc}
\end{small}
\end{center}
\end{table*}

\textbf{Tools:} To detect power calculations we used two tools, CONSORT-TM (RRID:SCR\_021051) and SciScore (RRID:SCR\_016251) (Table \ref{tab:power-tools}).

\textbf{Curation:} We performed manual validation of 132 (+100 controls) articles that were previously assessed by both automated screening tools and found to have at least one difference. Additionally, there were 100 control articles where the tools agreed, including articles where at least one tool detected a statement on power calculations and articles where neither tool identified a statement on power calculation.
Each article was assessed by a human curator (RS) to assess the presence or absence of a statement regarding a priori power calculation. Any sentence that indicated that an a priori power- or sample size calculation was performed (e.g., “a priori power calculation present”) was rated as “Yes”. The criteria did not assess whether the authors reported a minimum set of information that is required for an informative a priori power calculation (e.g. expected effect size, effect size justification, outcome of interest, Type I and II error). Statements on post hoc power calculation and other forms of power calculation that were not done a priori with the aim to estimate a required sample size were excluded, as were statements that no power or sample size calculation were conducted (“a priori power calculation not present”). 

\textbf{Results:}\\
\textbf{Function learned: (CONSORT-TM AND SciScore)\\
Percent same with 80\% data splits: 100\%}
Both tools reach comparable accuracies, while SciScore has a higher precision and CONSORT-TM has a higher recall (Tables \ref{tab:performance-power} and \ref{tab:agreement-power analysis}). The ensemble achieves the best accuracy and precision, although it is comparable to the other tools. In this case, the ensemble’s increase in accuracy is relatively modest, showing that ensemble approaches do not always lead to large increases in accuracy.

\subsubsection*{Software tools}
\begin{table*}
\caption{\textbf{List of tools for software tool detection.}}
\label{tab:software-tools}
\begin{center}
\begin{small}
\begin{sc}
\hspace*{-3cm}
\begin{scriptsize}
\begin{tabular}{c|c|cc}
\toprule
Tool category&Tool name&Searches for&Method of classification\\
\midrule
\multirow{2}{*}{Software tools} & SoftCite & \parbox{3cm}{Sentence(s) including software tools}&BERT encoding + linear classifier\\
\cmidrule{2-4}
& SciScore & \parbox{3cm}{Sentence(s) including software tools}&CRFs\\
\bottomrule
\end{tabular}
\end{scriptsize}
\end{sc}
\end{small}
\end{center}
\end{table*}

\begin{table*}
\caption{\textbf{Estimated tool performance statistics for software tool detection}. Statistics are estimated based on the gold standard test set of 1,500 papers (see Methods for details). The Ensemble tool uses a combination of the results from all tools to make a classification.}
\label{tab:performance-software}
\begin{center}
\begin{small}
\begin{sc}
\begin{tabular}{c|cccc}
\toprule
Tool & Accuracy & Precision & Recall & F1\\
\midrule
SciScore & 0.25 & 0.58 & 0.17 & 0.27 \\ 
SoftCite & \textbf{0.80} & \textbf{0.86} & \textbf{0.88} & \textbf{0.87} \\ 
\midrule 
Ensemble & \textbf{0.80} & \textbf{0.86} & \textbf{0.88} & \textbf{0.87}\\ 
\bottomrule
\end{tabular}
\end{sc}
\end{small}
\end{center}
\end{table*}

\begin{table*}
\caption{\textbf{Gwet's agreement among each tool and the gold standard (``true'') for software tool detection}.}
\label{tab:agreement-software}
\begin{center}
\begin{small}
\begin{sc}
\begin{tabular}{cccccc}
\toprule
 & SciScore & SoftCite & Ensemble & True\\
\midrule
SciScore & 1.0 & -0.91 & -0.91 & -0.50\\
SoftCite & & 1.0 & 1.0 & 0.70\\
Ensemble & & & 1.0 & 0.70\\
True & & & & 1.0\\
\bottomrule
\end{tabular}
\end{sc}
\end{small}
\end{center}
\end{table*}

\textbf{Tools:} SoftCite (RRID:SCR\_024411) and SciScore (RRID:SCR\_016251) were run to detect whether software was found in the paper (Table \ref{tab:software-tools}). For the analysis of SciScore’s software tool detector, we extracted all existing or suggested Research Resource IDentifiers (RRIDs) from SciScore’s “Key Resources” table and checked for the presence of any RRIDs that corresponded to software or databases as reported in the SciCrunch Registry, a cooperative project aiming to catalog scientific resources (Ozyurt et al. 2016). Software resources were queried for using the “additional resource type” field for any “software”, the file we generated on 28-June-2022 included 7,230 software tools. A live list is accessible via the scientific tools registry at \url{https://rrid.site/data/source/nlx_144509-1/search?q=%2A&l=&sort=desc&column=Mentions%20Count&sort=desc&filter[]=Resource%20Type:software}. The most commonly used software tools mentioned in papers are statistical tools, such as SPSS or R, and image manipulation tools, such as Photoshop or ImageJ. Both open and commercial tools are included. The list of databases was obtained from the same registry, querying the additional resource types field for “database” on 28-June-2022. This list included 2,970 databases.

\textbf{Curation:} We performed manual curation of 1498 mentions. The human curator (AN) systematically classified entities to be “software” or “not software”. Three columns were updated during the curation. The first column was the `Decision Column`, where each entity was rated as “yes”, “no”, or “it’s complicated”. The other two columns contained notes regarding the entity, whether it had an RRID or not, and some notes about the entity for “it’s complicated” cases. 

In the first pass of all the rows, the entity was given a “yes” if it was a software entity. There are two steps to come to that conclusion. The initial classification involves checking if the entity was a URL and, if so, determining its relevance to software based on redirection. For non-URL entities, a detailed classification process considers the presence of code, GitHub repositories, versions, packages and the ability to download the code. If it had code associated with it and had applications of software techniques, it was considered a software entity and marked as yes. 

The exclusion principle is also clearly followed by the curator to identify entities that are not software. The protocol provides specific exclusions, such as entities related solely to databases, search functionality web pages, core facilities, organizational names, and hardware components without associated code or software control. Additionally, if they are just the company names and do not have direct software/code associated with it is not considered a software. All these entities just are marked no in the decision column. There were a few entries which were ambiguous in terms of whether they are software or not which were marked as complicated. These entities were revisited as a group later in the second pass of all the rows to come up with a definite decision. 

\textbf{Results:}\\
\textbf{Function learned: (SoftCite)\\
Percent same with 80\% data splits: 88\%}

SoftCite has a significantly higher accuracy, precision, recall, and F1 compared to SciScore (Tables \ref{tab:performance-software} and \ref{tab:agreement-software}). A significant contributor to this difference is likely that SciScore only considers the methods section, while SoftCite looks at the entire text of the paper.

\subsubsection*{Open code}
\begin{table*}
\caption{\textbf{List of tools for open code detection.}}
\label{tab:opencode-tools}
\begin{center}
\begin{small}
\begin{sc}
\hspace*{-3cm}
\begin{scriptsize}
\begin{tabular}{c|c|cc}
\toprule
Tool category&Tool name&Searches for&Method of classification\\
\midrule
\multirow{2}{*}{Open code detection} & SciScore & \parbox{3cm}{SciScore
Code availability statement OR code identifier} & CRF OR regex\\
\cmidrule{2-4}
& ODDPub & \parbox{3cm}{Code identifier AND open access AND result replication possible AND machine readable AND not reused} & Regex\\
\bottomrule
\end{tabular}
\end{scriptsize}
\end{sc}
\end{small}
\end{center}
\end{table*}

\begin{table*}
\caption{\textbf{Estimated tool performance statistics for open code detection}. Statistics are estimated based on the gold standard test set of 1,500 papers (see Methods for details). The Ensemble tool uses a combination of the results from all tools to make a classification.}
\label{tab:performance-code}
\begin{center}
\begin{small}
\begin{sc}
\begin{tabular}{c|cccc}
\toprule
Tool & Accuracy & Precision & Recall & F1\\
\midrule
SciScore & 0.98 & 0.55 & 0.67 & 0.6 \\ 
ODDPub & \textbf{1.0} & \textbf{0.95} & \textbf{0.92} & \textbf{0.94} \\ 
\midrule 
Ensemble & \textbf{1.0} & \textbf{0.95} & \textbf{0.92} & \textbf{0.94} \\ 
\bottomrule
\end{tabular}
\end{sc}
\end{small}
\end{center}
\end{table*}

\begin{table*}
\caption{\textbf{Gwet's agreement among each tool and the gold standard (``true'') for open code}.}
\label{tab:agreement-open code}
\begin{center}
\begin{small}
\begin{sc}
\begin{tabular}{cccccc}
\toprule
 & SciScore & ODDPub & Ensemble & True\\
\midrule
SciScore & 1.0 & 0.97 & 0.97 & 0.98\\
ODDPub &   & 1.0 & 1.0 & 1.0\\
Ensemble &   &   & 1.0 & 1.0\\
True &   &   &   & 1.0\\
\bottomrule
\end{tabular}
\end{sc}
\end{small}
\end{center}
\end{table*}

\textbf{Tools:} ODDPub (\url{https://github.com/quest-bih/oddpub}, RRID:SCR\_018385) and SciScore (RRID:SCR\_016251) were run for each paper’s full text and methods section (Table \ref{tab:opencode-tools}). Results were stored in a SQLite database. ODDPub explicitly classifies papers as having open code. SciScore results were based on the presence of “Code Information” in the rigor table. 

\textbf{Curation:} We performed manual validation of 98 (+100 control) articles. The human curator (PE) broke papers into “yes” or “no” categories for the presence of open code. Each paper was first searched for the presence of the words “github”, “code”, and “available at”. Then, each sentence that contained those words was analyzed. We only counted papers as “open code” if they mentioned code that was written, at least in part, by the authors and deposited on a software repository. Cases where authors merely referenced another open source code repository were not counted. If the authors mentioned a repository that contained mostly data, PE viewed the repository and checked for the presence of any associated code to make the final decision. After analyzing all the sentences that contained the above keywords, the curator also examined the beginning and end of the paper’s main text to find any statements about open code, as many journals place this information in data and code availability statements. If the paper contained a statement about open code, the curator only counted the statement if it referenced a code repository. Statements like “No original code was generated for this study” were not counted as open code statements.

After running all tools, all cases where one or more tools disagreed over the presence of open data or code were analyzed manually to make a final determination about the presence of open data or code. Open data/code was declared present when 1) the authors explicitly stated where data/code could be found in the body text (not in a supplementary file), 2) the data/code was hosted on a open repository (“data is available upon request” was not counted), and 3) the data/code was the work of the authors and was released along with the paper (mentioning a third-party repository was not counted).

\textbf{Results:}\\
\textbf{Function learned: (ODDPub)\\
Percent same with 80\% data splits: 77\%}

Performance characteristics of all tools for open data detection on full texts are presented in Tables \ref{tab:performance-code} and \ref{tab:agreement-open code}. The ensemble method learned to just use the output from ODDPub for its prediction, meaning SciScore provided no additional information beyond what ODDPub already provides. While ODDPub and SciScore had generally very similar predictions (Table \ref{tab:agreement-open code}), we selected 10 random papers where the tools reported different results and analyzed differences to determine why the tools differed. For 9 (90\%) of these papers, SciScore reported open code while ODDPub did not. Among these, 2 (22\%) were due to differences in what the tools were designed to do (i.e. both tools were “correct” according to their own reported purposes), 2 (22\%) were due to misclassifications by ODDPub (i.e. the sentence was not picked up by ODDPub’s regexes), and 5 (56\%) were due to misclassifications by SciScore (i.e. SciScore’s conditional random field [CRF] model misclassified a sentence). For the one paper (10\%) where SciScore did not report open code while ODDPub did, SciScore incorrectly did not find an open code statement. Overall, 80\% of the difference between tools is attributable to the ability of the respective classification tools, with 75\% of this difference arising from incorrect classifications by a machine learning model. This indicates that model choice and data curation, and even the decision to use a machine learning approach, are more important than the exact definitions of open data/code used by the tool developers, even if such definitions are very different.

\subsubsection*{Contaminated cell lines}
\begin{table*}
\caption{\textbf{List of tools for contaminated cell line detection.}}
\label{tab:cellline-tools}
\begin{center}
\begin{small}
\begin{sc}
\hspace*{-3cm}
\begin{scriptsize}
\begin{tabular}{c|c|cc}
\toprule
Tool category&Tool name&Searches for&Method of classification\\
\midrule
\multirow{2}{*}{Contaminated cell lines} & PCLDetector & \parbox{3cm}{Contaminated cell lines} & Embedding + CNN\\
\cmidrule{2-4}
 & SciScore & \parbox{3cm}{SciScore
Detects cell line sentences, and RRIDs, if an RRID is present and the cell line contains a warning, the warning is displayed} & CRFs\\
\bottomrule
\end{tabular}
\end{scriptsize}
\end{sc}
\end{small}
\end{center}
\end{table*}

\begin{table*}
\caption{\textbf{Estimated tool performance statistics for problematic cell line detection}. Statistics are estimated based on the gold standard test set of 1,500 papers (see Methods for details). The Ensemble tool uses a combination of the results from all tools to make a classification.}
\label{tab:performance-cell lines}
\begin{center}
\begin{small}
\begin{sc}
\begin{tabular}{c|cccc}
\toprule
Tool & Accuracy & Precision & Recall & F1\\
\midrule
PCLDetector & \textbf{0.99} & 0.76 & \textbf{0.74} & \textbf{0.75}\\
SciScore & 0.98 & \textbf{1.0} & 0.36 & 0.53\\ 
\midrule 
Ensemble & \textbf{0.99} & 0.76 & \textbf{0.74} & \textbf{0.75}\\
\bottomrule
\end{tabular}
\end{sc}
\end{small}
\end{center}
\end{table*}

\begin{table*}
\caption{\textbf{Gwet's agreement among each tool and the gold standard (``true'') for problematic cell lines}.}
\label{tab:agreement-cell lines}
\begin{center}
\begin{small}
\begin{sc}
\begin{tabular}{cccccc}
\toprule
 & PCLDetector & SciScore & Ensemble & True\\
\midrule
PCLDetector & 1.0 & 0.97 & 1.0 & 0.99\\
SciScore &   & 1.0 & 0.97 & 0.98\\
Ensemble &   &   & 1.0 & 0.99\\
True &   &   &   & 1.0\\
\bottomrule
\end{tabular}
\end{sc}
\end{small}
\end{center}
\end{table*}

\textbf{Tools:} SciScore (RRID:SCR\_016251), and PCLDetector (N/A) were used to detect mentions of problematic cell lines (Table \ref{tab:cellline-tools}). SciScore aims at detecting any kind of cell lines, whereas PCLDetector is only designed to detect problematic cell lines tabulated by the International Cell Line Authentication Committee (ICLAC) or Cellosaurus (filtering for comments of type ‘problematic cell line’). PCLDetector is based on the ICLAC Register of Misidentified Cell Lines (Version 10) and a dump of Cellosaurus (Version 34.0). The article text is processed by the scispaCy library loaded with the ‘en\_ner\_jnlpba\_md’ named entity recognition model. As a result, entities of type DNA, CELL\_TYPE, CELL\_LINE, RNA, and PROTEIN are tagged. Then, the tool searches for the detected cell lines and cell types in the list of problematic cell lines (ICLAC or Cellosaurus). Both tools were run for each paper’s methods section.

\textbf{Curation:} Manual curation of each extracted cell line was done by human curators (CL, GC). The curator assumed that if a cell line was listed as problematic in ICLAC, Cellosaurus, or SciCrunch, it was indeed problematic.

Some cell lines are known to be cross-contaminated or otherwise misidentified but are still wrongly used and described in many scientific publications \cite{horbach2018changing, oste2024misspellings}. The use of RRIDs to identify cell lines has shown to be effective in reducing such detrimental usage \cite{babic2019incidences}. Nevertheless most cell line mentions do not feature RRID. This is why automatically screening scientific texts to specifically detect problematic cell lines is a way to increase the quality of published paper. 

\textbf{Results:}\\
\textbf{Function learned: (PCLDetector)\\
Percent same with 80\% data splits: 82\%}

After running all tools on the 1500 paper set, PCLDetector found 48 papers (3\%) with potentially problematic cell lines. However, SciScore only marked one paper as having problematic cell lines. While SciScore correctly identified two cell lines in this paper as problematic, PCLDetector only found one of these two cell lines. However, as this single match was not enough for a meaningful comparison, we decided to instead manually examine (CL, GC) all cell lines extracted by SciScore, and look for problems in the suggested cell line entry in SciCrunch. This is a natural step for users who would use the SciScore report, but is not integrated into SciScore due to false positive concerns. Using this procedure, SciScore found 14 papers (1\%) with potentially problematic cell lines.

Tables \ref{tab:performance-cell lines} and \ref{tab:agreement-cell lines} show the performance characteristics of each tool in their ability to detect problematic cell lines. As shown, PCLDetector is able to identify many more problematic cell lines than the modified SciScore, although it has a higher rate of false positives. As PCLDetector is specifically designed to search for problematic cell lines, it is unsurprising that it has a higher recall than a tool which is designed to be more conservative. We also see that the ensemble tool does not add performance over PCLDetector by itself.

\subsubsection*{Baseline table detection}
\begin{table*}
\caption{\textbf{Estimated tool performance statistics for baseline table detection}. Statistics are estimated based on the gold standard test set of 1,500 papers (see Methods for details). The Ensemble tool uses a combination of the results from all tools to make a classification.}
\label{tab:performance-btables}
\begin{center}
\begin{small}
\begin{sc}
\begin{tabular}{c|cccc}
\toprule
Tool & Accuracy & Precision & Recall & F1\\
\midrule
baseline & \textbf{0.80} & \textbf{0.88} & \textbf{0.77} & \textbf{0.82}\\
{[unnamed]} & 0.67 & 0.75 & 0.68 & 0.71\\
\bottomrule
\end{tabular}
\end{sc}
\end{small}
\end{center}
\end{table*}

\begin{figure}
\begin{center}
\centerline{\includegraphics[width=\textwidth]{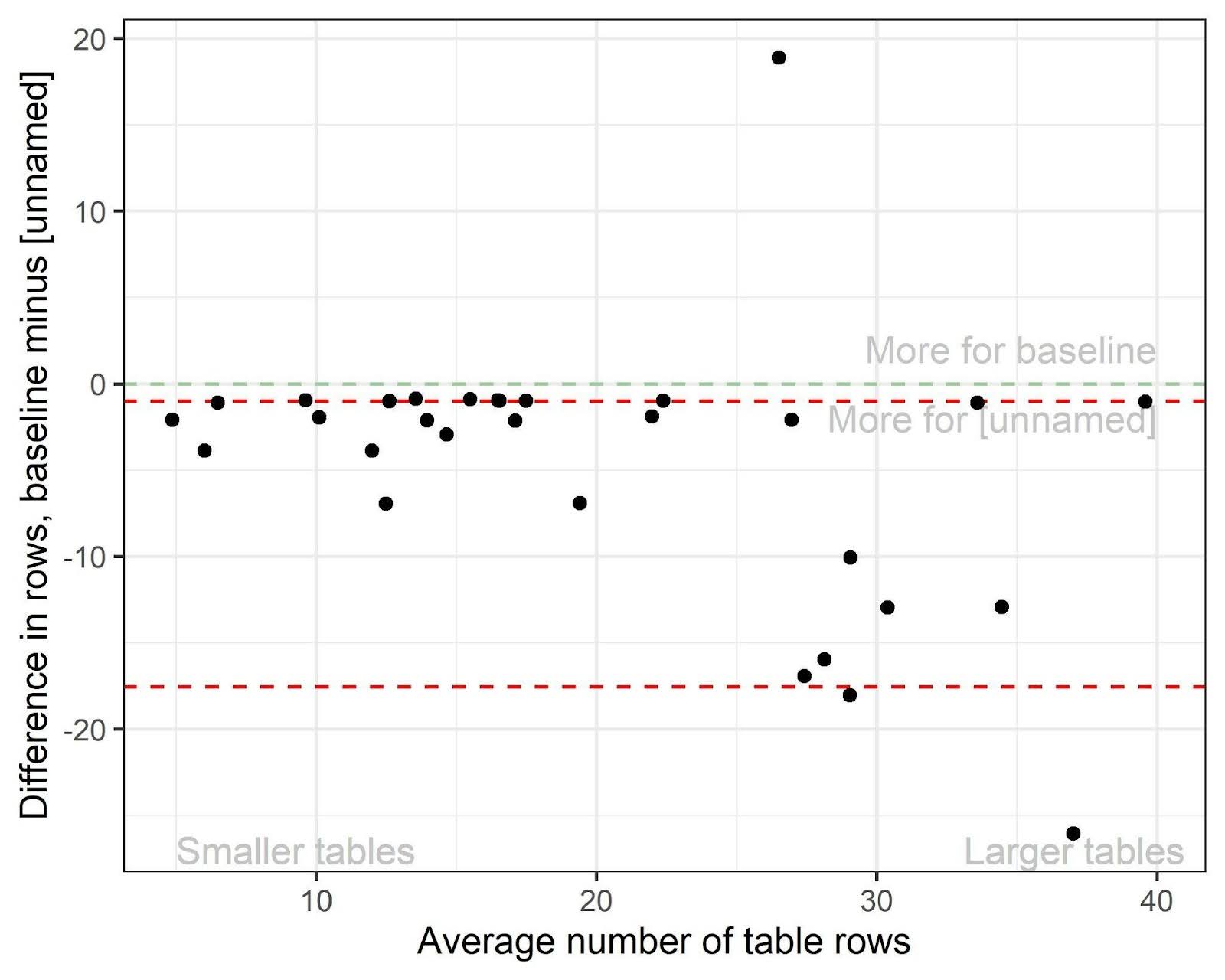}}
\caption{Bland–Altman plot of the number of rows in the baseline table extracted by the {[unnamed]} and baseline tools. The dotted red lines are the 95\% limits of agreement. The dotted green line at zero shows perfect agreement in table row numbers. }
\label{band-plot}
\end{center}
\end{figure}

\textbf{Tools:} We compared two tools that aim to extract the baseline table from randomized controlled trials, baseline (RRID:SCR\_025128) \cite{barnett2023automated} and {[unnamed]} \cite{vorland2021semi}. baseline searches the XML for key phrases that indicate a table comparing randomized groups at baseline using regex. {[unnamed]} searches the PDF for key phrases that indicate a table comparing groups at baseline using computer vision to extract tables and regex matching on the text. The CONSORT guidelines (\url{https://www.equator-network.org/reporting-guidelines/consort/}) for randomized trials recommend that papers reporting the results of a trial include a table displaying the characteristics of participants at baseline. The two tools were applied to a random sample of 100 papers that were sampled from \cite{carlisle2017data}. The code to extract the trials is available at \url{https://github.com/agbarnett/validation_trials}.

\textbf{Curation:} We did not use the 1500 paper test set because it contained many studies that were not trials and it was more useful to compare the tools’ ability to extract baseline tables in a set of papers that were likely to include many trials. Each paper was manually classified (AB, CJV) as including a baseline table or not, with group discussion used to resolve papers where it was unclear. 

\textbf{Results:} The ‘baseline’ tool did better on accuracy, precision and recall (Table \ref{tab:performance-btables}), however we should be cautious with these results as the sample size of papers is much smaller than the previous analyses, and different publication formats were used (i.e., XML vs. PDF). The main reasons that the XML-based algorithm excluded baseline tables were because: the table included follow-up comparisons, meaning the z-value distribution would no longer be expected to be centered on zero; the sample sizes could not be extracted; or because the tool did not classify the paper as a randomized trial. 

The baseline tool extracted tables using XML and the {[unnamed]} tool using the PDF. In addition to comparing whether the tools correctly extracted a baseline table or not, we also compared the size of the table extracted using a Bland–Altman plot of the number of rows. The plot shows that using the PDF meant the extracted tables were generally larger with more rows captured, and this was particularly noticeable for larger tables (Figure \ref{band-plot}). This likely occurred because some baseline tables were formatted as multiple tables in the XML although they appeared as one table when using the PDF. Smaller differences will have occurred because the XML algorithm only selects one row when there are two perfectly negatively correlated rows, e.g., summary statistics per randomized group on the number of men and women. 

\section*{Protocols}
\subsubsection*{Ensembling method (applicable to each comparison)}
\begin{enumerate}
    \item For each of the 1500 papers, store the binary output from each tool in an array, as well as the gold standard result from the test set
    \begin{enumerate}
        \item The array should have the shape (1500, number of tools)
    \end{enumerate}
    \item Train a support vector machine (SVM) classifier using sklearn on the full set of papers using the default hyperparameters
    \item Use the model to predict the results for the same 1500 paper set, which makes up the output for the “Ensemble” tool
    \item Get learned formula
    \begin{enumerate}
        \item Generate a truth table for all possible combinations of binary tool outputs, then use the SVM to predict the outcome
        \item Perform Boolean simplification on the truth table to generate a simplified equation
    \end{enumerate}
    \item Get \% same with subsets
    \begin{enumerate}
        \item Take input array and randomly sample half of the entries without replacement, and sample same indices in the output array
        \item Train SVM using this smaller subset, and perform (4) for each subset and count what percent of the models trained on random subsets match the model trained on the full dataset
    \end{enumerate}
\end{enumerate}

\subsubsection*{Inclusion / Exclusion Criteria}
Data preparation
\begin{enumerate}
    \item Download 1500 paper set
    \item Install and run pre-rob on the papers \begin{enumerate}
        \item Get data from the “exclusion” column in the output spreadsheet. If the value for a given paper is > 0.5, consider it as a positive, otherwise a negative.
    \end{enumerate}
    \item Repeat for Barzooka (\url{https://github.com/quest-bih/barzooka}) \begin{enumerate}
        \item Extract results from the 7th output column, which is a binary indication if any flowchart was found. 
    \end{enumerate}
    \item Repeat for SciScore \begin{enumerate}
        \item Get the “Inclusion and Exclusion Criteria” field in the Rigor Table from the outputted JSON, and consider any extracted sentence as a positive, and a “not required” or “not detected” as a negative.
    \end{enumerate}
    \item Aggregate data from the 3 tools into a spreadsheet with 2 columns \begin{enumerate}
        \item Take all papers where at least one tool disagreed on the binary presence of inclusion/exclusion criteria, and make a row for each
        \item First column in spreadsheet contains a link to the paper
        \item Second column contains a sentence extracted from a randomly selected tool. If it found inclusion/exclusion criteria, show the sentence, otherwise show no sentence. Do not show anything for Barzooka.
    \end{enumerate}
\end{enumerate}

Curation
\begin{enumerate}
    \item Add 3 additional columns \begin{enumerate}
        \item First column is a “Decision” column that can take on the value “yes”, “no”, or “complicated”
        \item Second column and third columns contain notes
    \end{enumerate}
    \item Conduct first pass on all rows \begin{enumerate}
        \item For each row, check if “inclusion” or “exclusion” are present in the extracted sentence and the sentence designates a group or subject that is included/excluded. If yes, mark the decision as “yes”. If sentence mentions including/excluding an experimental variable that is not a subject, then mark the decision as “no”. If sentence does not clearly fall into these categories, mark as  “complicated”. If there is no sentence associated with the row, skip.
    \end{enumerate}
    \item Conduct second pass on rows marked “complicated” and those without sentences \begin{enumerate}
        \item For each row, open up linked paper and look for any sentence or diagram in the manuscript that would indicate inclusion/exclusion. If a clear sentence could be found, paste it into the document in the notes column. Using information from the full paper, make a decision on whether inclusion or exclusion was present (removing the “complicated” designation).
    \end{enumerate}
\end{enumerate}

\subsubsection*{Blinding}
Data preparation
\begin{enumerate}
    \item Download 1500 paper set
    \item Install and run pre-rob on the papers \begin{enumerate}
        \item Get data from the “blind” column in the output spreadsheet. If the value for a given paper is > 0.5, consider it as a positive, otherwise a negative.
    \end{enumerate}
    \item Repeat for SciScore \begin{enumerate}
        \item Get the “Blinding” field in the Rigor Table from the outputted JSON, and consider any extracted sentence as a positive, and a “not required” or “not detected” as a negative.
    \end{enumerate}
    \item Repeat for CONSORT-TM \begin{enumerate}
        \item Get data from the “BLINDING” column in the output spreadsheet. Consider a paper a positive if any sentence within that paper has “TRUE” in the “BLINDING” column.
    \end{enumerate}
    \item Aggregate data from the 3 tools into a spreadsheet with 2 columns \begin{enumerate}
        \item Take all papers where at least one tool disagreed on the binary presence of blinding, and make a row for each
        \item First column in spreadsheet contains a link to the paper
        \item Second column contains a sentence extracted from a randomly selected tool. If it found blinding, show the sentence, otherwise show no sentence.
    \end{enumerate}
\end{enumerate}

\subsubsection*{Randomization}
Data preparation
\begin{enumerate}
    \item Download 1500 paper set
    \item Install and run pre-rob on the papers \begin{enumerate}
        \item Get data from the “random” column in the output spreadsheet. If the value for a given paper is > 0.5, consider it as a positive, otherwise a negative.
    \end{enumerate}
    \item Repeat for SciScore \begin{enumerate}
        \item Get the “Randomization” field in the Rigor Table from the outputted JSON, and consider any extracted sentence as a positive, and a “not required” or “not detected” as a negative.
    \end{enumerate}
    \item Repeat for CONSORT-TM \begin{enumerate}
        \item Get data from the “RANDOMIZATION” column in the output spreadsheet. Consider a paper a positive if any sentence within that paper has “TRUE” in the “RANDOMIZATION” column.
    \end{enumerate}
    \item Aggregate data from the 3 tools into a spreadsheet with 2 columns \begin{enumerate}
        \item Take all papers where at least one tool disagreed on the binary presence of randomization, and make a row for each
        \item First column in spreadsheet contains a link to the paper
        \item Second column contains a sentence extracted from a randomly selected tool. If it found blinding, show the sentence, otherwise show no sentence.
    \end{enumerate}
\end{enumerate}

\subsubsection*{Power calculations}
Data preparation
\begin{enumerate}
    \item Download 1500 paper set
    \item Install and run SciScore on the papers \begin{enumerate}
        \item Get the “Power Analysis” field in the Rigor Table from the outputted JSON, and consider any extracted sentence as a positive, and a “not required” or “not detected” as a negative.
    \end{enumerate}
    \item Repeat for CONSORT-TM \begin{enumerate}
        \item Get data from the “SAMPLE\_SIZE\_CALCULATION” column in the output spreadsheet. Consider a paper positive if any sentence within that paper has “TRUE” in the “SAMPLE\_SIZE\_CALCULATION” column.
    \end{enumerate}
    \item Aggregate data from the 3 tools into a spreadsheet with 2 columns \begin{enumerate}
        \item Take all papers where at least one tool disagreed on the binary presence of power calculations, and make a row for each
        \item First column in spreadsheet contains a link to the paper
        \item Second column contains a sentence extracted from a randomly selected tool. If it found power calculations, show the sentence, otherwise show no sentence.
    \end{enumerate}
\end{enumerate}

\subsubsection*{Open code detection}
Data preparation
\begin{enumerate}
    \item Download 1500 paper set
    \item Install and run SciScore on the papers \begin{enumerate}
        \item Extract the “Code Information” from the SciScore Rigor table
    \end{enumerate}
    \item Repeat for ODDPub \begin{enumerate}
        \item Get the column for open code in the output spreadsheet
    \end{enumerate}
    \item Aggregate data from the 3 tools into a spreadsheet with 2 columns \begin{enumerate}
        \item Take all papers where at least one tool disagreed on the binary presence of open code, and make a row for each
        \item First column in spreadsheet contains a link to the paper
        \item Second column contains a sentence extracted from a randomly selected tool. If it found open code, show the sentence, otherwise show no sentence.
    \end{enumerate}
\end{enumerate}

\subsection*{Software tools}
\begin{enumerate}
    \item Download 1500 paper set
    \item Install and run SciScore on the papers \begin{enumerate}
        \item Extract the “Software Tools” objects from the Key Resources table
    \end{enumerate}
    \item Repeat for SoftCite
    \item Initial review decision column update:\begin{enumerate}
        \item Yes: If the entity is confirmed as software.
        \item No: If the entity is not software.
        \item Complicated: If the classification of the entity is ambiguous and needs further review.
    \end{enumerate}
    \item Classification Criteria: \begin{enumerate}
        \item For URL Entities: Check if the URL is active and redirects to relevant software-related content.
        \item For Non-URL Entities: Examine the presence of code, GitHub repositories, software versions, packages, and download options. Then confirm applications of software techniques or development practices.
    \end{enumerate}
    \item Exclusion Criteria: \begin{enumerate}
        \item Entities solely related to databases, search functionalities, core facilities, organizational names, and hardware components without software or code integration.
        \item Entities representing company names without direct involvement in software development or code provision.
    \end{enumerate}
    \item Additional Columns: \begin{enumerate}
        \item RRID Presence: Indicate whether the entity has an RRID (Research Resource Identifier).
        \item Notes: Provide detailed notes for entities marked as "Complicated," including the reason for ambiguity and any additional information relevant for the second review.
    \end{enumerate}
    \item Second Review Pass: \begin{enumerate}
        \item Revisit "Complicated" entities as a group to resolve ambiguities and make final classifications.
        \item Update the Decision Column based on collective assessments and additional information gathered.
    \end{enumerate}
\end{enumerate}

\subsubsection*{Registration reporting}
\begin{enumerate}
    \item Download 1500 paper set
    \item Install and run SciScore on the papers \begin{enumerate}
        \item Extract any mention results from the “Protocol Information” section in the Rigor Table
    \end{enumerate}
    \item Install and run TRNscreener on the papers \begin{enumerate}
        \item Run file trial\_identifier\_search.R on the folder of full texts, and extract identifiers from the 3rd column of the output
    \end{enumerate}
    \item Install and run ctregistries on the papers
    \item Search for “NCT” in the paper full texts \begin{enumerate}
        \item No regex was used, only searching for the presence of “NCT”
    \end{enumerate}
\end{enumerate}

\subsubsection*{Contaminated cell lines}
\begin{enumerate}
    \item Download 1500 paper set
    \item Install and run SciScore on the papers \begin{enumerate}
        \item Extract the “Cell Line Authenticity” field from the Rigor Table
    \end{enumerate}
    \item Repeat for PCLDetector
\end{enumerate}

\section*{Abbreviations}
XML - extensible markup language
PMCID - PubMed Central ID
RRID - research resource identifiers
TP, TN, FP, FN - true positive, true negative, false positive, false negative

\end{document}